\documentclass[12pt]{article}
\usepackage{bm}
\usepackage{a4wide,graphicx}
\usepackage{cite}
\begin{document}


\title{Open charm in nuclear matter at finite temperature}

\author{L. Tol\'os $^{1}$, A. Ramos$^2$ and T. Mizutani$^{3}$\\
$^1$ Frankfurt Institute for Advanced Studies.
J.W. Goethe-Universit\"at,\\
Ruth-Moufang-Str. 1, 60438 Frankfurt am Main, Germany\\
$^2$ Departament d'Estructura i Constituents de la Mat\`eria\\
Universitat de Barcelona,
Diagonal 647, 08028 Barcelona, Spain \\
$^3$ Department of Physics, Virginia Polytechnic Institute and State
University \\
Blacksburg, VA 24061, USA}

\date{\today}

\maketitle
\begin{abstract}

We study the properties of open-charm mesons ($D$ and $\bar {D}$)
in nuclear matter at finite temperature within a self-consistent
coupled-channel approach.  The meson-baryon  interactions are
adopted from a type of broken SU(4) $s$-wave Tomozawa-Weinberg terms
supplemented by an attractive scalar-isoscalar interaction. The
in-medium solution at finite temperature incorporates Pauli
blocking effects, mean-field binding on all the baryons involved,
and $\pi$ and open-charm meson self-energies in a self-consistent
manner. In the $DN$ sector, the $\Lambda_c$ and $\Sigma_c$
resonances, generated dynamically at 2593 MeV and 2770 MeV in free
space, remain close to their free-space position while acquiring a
remarkable width due to the thermal smearing of Pauli blocking as
well as from the nuclear matter density effects. As a result, the
$D$ meson spectral density shows a single pronounced peak for 
energies close to the $D$ meson free-space mass that
broadens with increasing matter density with an extended tail
particularly towards lower energies. The $\bar D$ potential shows a moderate repulsive
behavior coming from the dominant $I=1$ 
contribution of the ${\bar D}N$ interaction.  The
low-density theorem is, however, not a good approximation for the
$\bar D$ self-energy in spite of the absence of resonance-hole 
contributions close to threshold in this case. 
We speculate the possibility of $D$-mesic nuclei as
well as discuss some consequences for the $J/\Psi$ suppression in
heavy-ion collisions, in particular for the future CBM experiment
at FAIR.

\end{abstract}
\vskip 0.5 cm




\section{Introduction}
\label{sec:intro}

The interest in the open and hidden charmed mesons within the context of
relativistic nucleus-nucleus collisions was triggered about 20 years ago. More
 specifically, the suppression of the $J/\Psi$ production  was predicted
as a rather clear signature of the formation of quark-gluon plasma (QGP) in
ultra-relativistic central  nucleus-nucleus collisions in Ref.~\cite{MAT86}.
According to its authors, a Debye-type color screening  in
the gluon exchanges blocks the formation of charmonium ($c \bar c$) bound
states. Then, starting about 10 years later,   the NA50 and NA60 Collaborations
(see for example \cite{GON96, RAM98, ABR00, FOE06}, in the CERN SPS fixed
target  experiments) actually claimed to have observed such a suppression in
Pb + Pb collisions   at $\approx 160$A GeV. Upon the start of the Brookhaven
RHIC heavy nucleus collider a little more than 6 years ago with, say, central
collision of Au + Au at  $\sqrt{s_{NN}}= 200$ GeV, a new set of exciting results
has gradually come out, such as an apparent  energy independence of the
$J/\Psi$ suppression \cite{QMA06} as compared with the SPS result with $\sqrt{s_{NN}}=17$ 
GeV. Firmly establishing the origin of this charmonium suppression as
due to  the formation of the QGP appears to need more careful analyses of the
data. However, if formed in such ultra-relativistic collisions, the QGP would
correspond to the one with a rather high temperature: $T > T_c$, where the
critical temperature extracted from recent QCD lattice simulation \cite{AOK06} is
$T_c \approx 175$ MeV with a low-baryon number
density $\rho_B$ (or chemical potential $\mu_B$), supposedly similar to the
situation during the initial Big-Bang period. This part of the $T-\mu_B$ phase
diagram of the hadronic/partonic {\it matter} will continue to be the major
subject of further intense activities at RHIC as well as at the CERN LHC
facility (expected at $\sqrt{s_{NN}} > 5$ TeV for Pb+Pb).

Equally interesting as well as important is the somewhat complementary region
of the phase diagram which is characterized by a moderate  temperature but with
large $\rho_B$.  According to recent lattice simulations (see for example
\cite{FOD04}), here  a highly compressed hadronic matter gets transformed into
a dense partonic matter (or strongly interacting QGP)  where the boundary of
the two phases  is characterized by  a first-order phase transition, as
opposed to the above-mentioned RHIC-type hadron $\leftrightarrow$ QGP
transition which the calculation has found as a smooth cross over (see also a
so-called Polyakov loop extended NJL (PNJL) approach to the subject
\cite{RAT06}). The CBM (Compressed Baryon Matter) experiment of the FAIR
project at GSI aims at investigating an important portion of
this  moderate $T$ and large $\mu_B$ part of the phase diagram by a high-intensity beam of, for example, uranium nuclei of  up to $35$A GeV which overlaps 
with the SPS energy. In this way
one may expect to study possible modifications of the properties of
various mesons in dense baryonic matter. In
particular, since the charmed mesons  produced at FAIR will not be at  high
energies but could be close to threshold, their medium modification may
be significative.  This should apply equally to the production of
open-charm mesons such as $D$ and $\bar D$ as well as  to hidden charmed mesons:
charmonia.  For the latter,  one will be able to study the
possible suppression and regeneration of the $J/\Psi$  at moderate energies
by mechanisms that may be of conventional hadronic origin,  or due to
deconfinement but different from the high  $T$(QGP) color screening
scheme  proposed in \cite{MAT86} and being sought by the far higher
energy RHIC accelerator, as stated earlier.

 Our present interest is in
relativistic heavy-ion collisions which  fit into some part of the domain of
the phase diagram    covered by the FAIR project.  In particular, we would like
to focus on hadronic approaches to the in-medium modification of the $D
(\bar D)$ mesons which may

\noindent (i) enter the  explanation of the possible $J/\Psi$ suppression in
relativistic nucleus-nucleus collisions, see for example Ref.~\cite{TSU01},  with
special interest in the FAIR energies.

\noindent (ii) provide a theoretical support for an anticipated open-charm enhancement
again within the FAIR energies \cite{CAS01}, an issue that was
triggered by the NA50 Collaboration, but was not recognized by the NA60
result, see \cite{SCO05}.

\noindent (iii) infer possible $D^0, D^-, \bar  {D^o}$ bound states in heavy
nuclei such as $Pb$ \cite{TSU99}.

Here we should stress that all these interpretations/predictions
are based upon the possible attraction felt by the  $D (\bar D)$
mesons which could lead to their mass reduction in the nuclear
medium \cite{TSU99, SIB99, HAY00, MIS104}.  For example, within
these mostly hadronic pictures, the $J/\Psi$ absorption by
collision with nucleons and mesons was  suggested to take place at
little or no extra cost of energy due to the lowering of the
threshold for $D \bar D$ pairs, facilitating reactions of the
$J/\Psi$ with comoving mesons, such as $J/\Psi \pi \to  D \bar D$.
In a similar manner, processes such as  $J/\Psi N \to \bar D Y_c$
where $Y_c$ is one of the charmed baryons may proceed more easily.
A critical and detailed review on these {\it mean field}
approaches was made in \cite{MIZ06} which eventually  points to
the necessity of performing  a coupled-channels meson-baryon
scattering in nuclear medium  due to strong coupling among the
$DN$ and other meson-baryon  channels with same quantum numbers.
Hence,  in the present article, we pursue a coupled-channel study
on the spectral properties of the open-charm  $D$ and $\bar D$
mesons in nuclear matter at finite temperatures.    In this
regard,  we want to remark that  kinetic equilibrium  assumed in
the corresponding heavy-ion reactions to introduce a well-defined
temperature may yet to be firmly established,  although, as
pointed out in  \cite{CAS01},  non-equilibrium transport equation
methods appear to support the thermalization picture \cite{GAZ991,
GAZ992, GAL00}  at SPS  energies.  To set the basis, we should
mention here earlier prototypes to the present study. First, a
coupled-channel approach based on a set of separable meson-baryon
interactions was adopted with an underlying $SU(3)$ symmetry among
the $u-, d-, c-$ quarks (thus excluding the strangeness related
channels). After model parameters were fixed to reproduce the
position and width of the $\Lambda_c(2593)$, it was  applied to
study the in-medium  spectral function of the $D$ meson for a zero
temperature nuclear matter environment\cite{TOL04}. This was later
extended  to finite temperatures \cite{TOL06}. Next, based upon a
$SU(4)$ scheme broken by the masses of the exchanged vector mesons
(the charmed ones in particular), hence including also the
channels with strangeness \cite{HOF05}, an effective meson-baryon
interaction, of lowest order in both chiral and heavy-quark
symmetries, was introduced to study the $D$ mesons in nuclear
matter at zero temperature \cite{LUT06, MIZ06}.

To continue and complete this sequence, in the present article we are
extending the result of Ref. \cite{MIZ06}  to study the $D$ and $\bar D$ mesons
in nuclear matter of up to twice the normal density, and having a temperature
from zero up to $150$ MeV.  A couple of extra additions in the present work
include: (a)  an explicit consideration of the nuclear  mean-field binding
effect on all the baryons involved in the coupled channels, inclusive of
strange and charmed, and (b) the study of the in-medium $\bar D$  (note
that ~\cite{LUT06} also looked at this aspect). Equipped with those
tools and ingredients, we go beyond the result of Refs.~\cite{TSU99, SIB99,
HAY00, MIS104} and try to determine the continuous in-medium spectral (or {\it
mass}) distribution of $D$ and $\bar D$, and show that the former deviates
significantly from the delta-function type spike at the free-space mass or at
the value shifted to elsewhere. Also we obtain optical potentials for  $D$ and
$\bar D$. These are important ameliorations in view of  the points (i)-(iii)
stated above.


 The organization of the present article goes as follows: in
Sect.~\ref{sec:Form} we develop the formalism and ingredients on which the
calculation in the present work is based. Sect.~\ref{sec:Resul} is devoted to
the presentation and discussion of the results. Finally, in
Sect.~\ref{sec:Conclusion}  we draw our conclusions and give
final remarks pertaining
to the present and future works.


\section{Open-charm mesons in nuclear matter \\ at finite temperature}
\label{sec:Form}

Our objective in this section is to obtain the $D$ and $\bar D$ in-medium
self-energy  by solving the corresponding multi-channel $T$ matrix
equation. Then, the obtained self-energy is used to find the  $D (\bar
D)$ spectral function in an iso-symmetric nuclear matter at finite temperature.
This is done by extending the  procedure found in Ref. \cite{MIZ06} to a
non-zero temperature environment by the procedure adopted in
Ref.~\cite{TOL06}.  As mentioned in Sect.~\ref{sec:intro},  we shall also
introduce the binding effect to all the baryons involved by the nuclear matter
mean field.


\subsection{Coupled meson-baryon channels in free space}

The first step towards our goal is to obtain the free space $T$-matrices for
the coupled meson-baryon system  involving $DN (\bar D N)$. We shall briefly
summarize it as discussed in Ref. \cite{MIZ06}. These matrices follow the
standard  multi-channel scattering (integral) equation, \begin{equation} T =  V
+ V G T, \end{equation} \noindent where $V$ is  a symmetric matrix consisting
of a set of meson-baryon transition interactions (potentials). They are
obtained from  the tree level $s$-wave contribution to the meson-baryon
scattering, and will be specified later. As shown in Ref. \cite{Oller,OSE98},  the
kernel of the equation  for $s$-wave interaction can be factorized in the
on-mass-shell ansatz, leaving the  four-momentum integration only in the
two-particle meson-baryon  propagators. These quantities often called loop
functions form a diagonal matrix $G$. They are divergent and  thus need to be
regularized. In the present work the cut-off method is adopted as it is more
appropriate than dimensional methods when dealing with  particles in a  medium,
as done in Refs.~\cite{RAM00,TOL2006} where it is applied in the study of
$\bar K$ in nuclear matter.

 The consequence of the on-shell ansatz is a set of linear algebraic equations whose solution now reads,
\begin{equation}
T = [1 - V G ]^{-1} V,
\label{eq:BSalgeb}
\end{equation}
which is practically equivalent to the so-called $N/D$ method \cite{Oller}.

The meson-baryon transition interaction $V$ is characterized  here by the
channel quantum numbers, charm $C$, strangeness $S$  and isospin $I$.  We
implicitly fix the first two quantum numbers and label $V$ explicitly by $I$.
For each ($C, S$) fixed, the coupled-channel elements of the
symmetric matrix $V$ of a given isospin $I$ are specified as $V^I_{ij}$
 for  a transition $i \leftrightarrow j$.  For
$(C=1, S=0)$, $i$ and $j$ run through

\noindent $\pi \Sigma_c$(2589), $DN$(2810),$\eta \Lambda_c$(2835),
$\ K \Xi_c$(2960), $\ K \Xi_c^\prime$(3071),
$D_s \Lambda$(3085) and $\eta^\prime \Lambda_c$(3245)

\noindent for the $I=0$ sector, and

\noindent $\pi \Lambda_c$(2425), $\pi \Sigma_c$(2589), $DN$(2810),
$\ K \Xi_c$(2960), $\eta \Sigma_c$(3005), $K \Xi^\prime_c$(3071),
$D_s \Sigma$(3160) and $\eta^\prime \Sigma_c$(3415)

\noindent for the $I=1$ sector. In the case ($C=-1,S=0$), there is only a
single channel, $\bar D N(2810)$, for each isospin $I=0$ and $I=1$ value. We
note that, in the above description, the value in the parentheses denotes the
channel threshold in MeV.

The concrete form for the matrix elements of $V$ comes from the $s$-wave Tomozawa-Weinberg
(T-W) term  as the zero-range limit of the lowest-order interaction of the SU(4) pseudoscalar meson and $1/2^+$ ground-state baryon multiplets based on the universal
vector-meson coupling hypothesis equipped with the extended KSFR condition (see Ref. \cite{MIZ06} for details):
\begin{equation}
V^I_{ij}(\sqrt{s})= -\frac{\kappa C_{ij}}{4f^2}(2\sqrt{s}-M_i-M_j)
\left(\frac{M_i+E_i}{2M_i}\right)^{1/2}\left(\frac{M_j+E_j}{2M_j}\right)^{1/2}
\ .
\label{eq:TWint}
\end{equation}
The coupling strength $C_{ij}$  derives from the $SU(4)$ symmetry
for the $i \leftrightarrow j$ transition, $\sqrt{s}$ is the center
of mass energy, $f=1.15 \ f_{\pi}$, where the value of $f$ has been adopted from
Ref.~\cite{OSE98}, and $M_i$ and $M_j$ as well as $E_i$ and $E_j$ are
the masses and energies  of baryons in channels $i$ and $j$,
respectively.  The breaking of the SU(4) symmetry in the T-W interaction 
through the physical hadron masses is mostly eminent in the reduction factor 
$\kappa$ which is $unity$ for
transitions $i \leftrightarrow j$, driven by uncharmed
vector-meson exchanges ($\rho,\ \omega,\ \phi, \ K^*$) but
is equal to $\kappa_c=(\bar m_V/\bar m_V^c)^2\approx 1/4$  for
charmed vector-meson exchanges such as $D^*$ and $D_s^*$, where
$\bar m_V$($\bar m_V^c$) is the mass of the typical  uncharmed
(charmed) exchanged vector meson.   The transition coefficients
$\tilde C_{ij}\equiv \kappa C_{ij}$, which are symmetric with
respect to the indices, are listed in Tables~I and II of Ref.
\cite{MIZ06} for the $DN$ sector.  For the $\bar D N$ system,
$\tilde C_{\bar D N \rightarrow \bar D N }^{I=0}=0$ and $\tilde
C_{\bar D N \rightarrow \bar D N}^{I=1}=-2$. The reader is
reminded of the same situation for the $KN$ coupling strengths due
to $SU(3)$ symmetry.  The T-W vector interaction is supplemented
by  a scalar-isoscalar attraction, which turns out to be important
in kaon condensate studies (see Ref. \cite{MIS104,MAR05}). The
$s$-wave projection of this interaction is equal to
\begin{equation}
V_{\Sigma}(\sqrt{s})=-\frac{\Sigma_{DN}}{f_D^2}\left(\frac{M_N+E}{2M_N}\right)\ ,
\label{eq:sigma}
\end{equation}
and it is independent of the $C$ and $I$ specification in the
present context. Here, $f_D$ is the $D(\bar D)$ meson weak decay
constant, and $\Sigma_{DN}$ is the strength of this interaction.
Note that for simplicity we introduce this  only in the diagonal
$D(\bar D)N$ interaction. The most recent determination of
$f_D=157$ MeV may be found in \cite{ART05}. 
As for the value of $\Sigma_{DN}$, we
simply  follow  what a QCD sum-rule \cite{HAY00} and a nuclear
mean-field approach of \cite{MIS104} have suggested, and estimate
it conservatively as $\Sigma_{DN} \approx 2000$ MeV. Because we can only
aim at qualitative estimates, we set $f_D \sim 200$ MeV and 
determine the strength of the scalar-isoscalar interaction to be
$\Sigma=\Sigma_{DN}/f_D^2=0.05$ MeV$^{-1}$.
We also accommodate the case where no such attraction is added, hence
$\Sigma_{DN}=0$. Based upon the above interactions, the
multi-channel transition $T$ matrices are solved such that the
momentum  cut-off $\Lambda$ is fixed to reproduce the position and
the width of the $I=0$ $\Lambda_c(2593)$ resonance. The parameters
for the  two cases adopted from ~\cite{MIZ06} are used in the
present work, model A: $f=1.15f_{\pi},\
\Sigma=0.05$  MeV$^{-1}$, $\Lambda=727$ MeV, and
model B: $f=1.15f_{\pi}$, $\Sigma=0$ MeV$^{-1}$,
$\Lambda=787$  MeV.
 These two model interactions produce a
resonance in $I=1$ channel whose position and width are: $2770$
MeV, $~20$ MeV (model A),  and $2795$ MeV, $~20$ MeV (model B),
respectively, close to the nominal $\Sigma_c(2800)$ \cite{pdg}.

\subsection{Coupled meson-baryon interaction in finite-temperature nuclear matter with mean-field binding}


The properties of the $D(\bar D)$ mesons in nuclear matter at
finite temperature and with mean-field binding  are obtained by
incorporating the corresponding medium modifications in the loop
function matrix $G$ only.  That is,  we assume that the
interaction matrix $V$ stays unchanged in medium.   In
Eq.~(\ref{eq:TWint}), the two square rooted factors  come from the
baryon spinor normalization, hence should stay more or less the
same.  The finite-temperature nuclear mean-field binding makes the
in-medium baryon masses $M^*_i(T)$ to be shifted from their free
space values $M_i$, so the second factor in the  expression should
be different in medium.  We nevertheless retain the vacuum mass
values in this interaction as we have estimated the effect to  be
at most a few percent in magnitude. For the same reason the
scalar-isoscalar interaction Eq.~(\ref{eq:sigma}) remains
unchanged. By implicitly understanding that $G$ and $T$ are now to
be interpreted as medium modified, the same multi-channel
algebraic equation Eq.~(\ref{eq:BSalgeb}) is to be solved.

Let us now describe in detail the medium and temperature
modifications. First, the baryons in the coupled channels, namely
the nucleon, $\Lambda, \Sigma, \Lambda_c$ and $\Sigma_c$, change both
their mass and energy-momentum relations due to the
finite-temperature mean-field binding effect.  We have adopted a
temperature dependent Walecka-type $\sigma -\omega$ model to
account for this change, see for example Ref.~\cite{KAP-GALE}.
Within this model, the nucleon energy spectrum in mean-field
approximation is obtained from
\begin{eqnarray}
E_N(\vec{p},T)=\sqrt{\vec{p}\,^2+M_N^*(T)^2}+\Sigma^v \ ,
\end{eqnarray}
with the vector potential $\Sigma^v$ and the effective mass $M_N^*(T)$ given  by
\begin{eqnarray}
\Sigma^v&=&\left(\frac{g_v}{m_v}\right)^2 \rho    \nonumber \\
M_N^*(T)&=&M_N-\Sigma^s, ~~~~~~~~~{\rm with}~\Sigma^s=
\left(\frac{g_s}{m_s}\right)^2 \rho_s \ ,
\end{eqnarray}
where $m_s$ and
$m_v$ are the meson masses, while $g_s$ and $g_v$ are the density dependent scalar and vector coupling constants, respectively. These constants are obtained by reproducing the energy per particle of symmetric nuclear matter at $T=0$ coming from a Dirac-Brueckner-Hartree-Fock
 calculation (see  Table 10.9
of Ref.~\cite{Machleidt}).
The {\it  ordinary} nuclear matter (Lorentz) vector density ($\rho$) and (Lorentz) scalar
density ($\rho_s$) are obtained from the
corresponding vector
 ($n(\vec p, T)$) and scalar  ($n_s(\vec p, T)$) density distributions  defined in terms of the Fermi-Dirac function as
\begin{equation}
n(\vec{p}, T)=\frac{1}{1+\exp{\left [(E_N(\vec{p}, T)-\mu)/T\right ]}},\ \ n_s(\vec{p}, T)=\frac{M^*(T)n(\vec p,T)}{\sqrt{\vec{p}\,^2+M^*(T)^2}},
\label{eq:density-dist}
\end{equation}
by momentum integration, namely
\begin{equation}
\rho= \frac{4}{(2\pi)^3} \int d^3p \ n(\vec{p},T)
\label{eq:density}
\end{equation}
(and similarly for $\rho_s$).
As may be clear from the above development, $E_N(\vec{p}, T),
M_N^*(T)$ and the chemical potential $\mu$ are obtained
simultaneously and self-consistently for given $\rho$ and for the
corresponding values of $g_s$ and $g_v$. The values of the nucleon
scalar and vector potentials in nuclear matter at  $\rho_0=0.16$ fm$^{-3}$ and at $T=0$ are $\Sigma^s= 356$ MeV
and $\Sigma^v= 278$ MeV.

The hyperon ($Y$) as well as the charmed baryon ($Y_c$) masses and energy spectra can be
easily inferred from those for the nucleon as
\begin{eqnarray}
E_{Y_{(c)}}(\vec{p},T)=\sqrt{\vec{p}\,^2+M_{Y_{(c)}}^*(T)^2}+\Sigma_{Y_{(c)}}^v \ ,
\end{eqnarray}
where
\begin{eqnarray}
\Sigma_{Y_{(c)}}^v&=&\frac{2}{3} \left(\frac{g_v}{m_v}\right)^2 \rho= \frac{2}{3}\Sigma^v  \nonumber \ ,\\
M_{Y_{(c)}}^*(T)&=&M_{Y_{(c)}}-\Sigma_{Y_{(c)}}^s=M_{Y_{(c)}}-\frac{2}{3}\left(\frac{g_s}{m_s}\right)^2 \rho_s \nonumber \\
&=& M_{Y_{(c)}}-\frac{2}{3}(M_N-M_N^*(T)) \ .
\end{eqnarray}
Here we have assumed that the $\sigma$ and $\omega$ fields only couple to the
$u$ and $d$ quarks, as in Ref.~\cite{TSU03},
so the scalar and vector coupling constants for hyperons and charmed baryons are:
\begin{eqnarray}
g_v^{Y_{(c)}}=\frac{2}{3}g_v , \hspace{1cm} g_s^{Y_{(c)}}=\frac{2}{3}g_s .
\end{eqnarray}
We note that the quark-meson coupling (QMC) calculations of
Ref.~\cite{TSU03}, performed at $T=0$, obtained a somewhat
smaller scalar potential (about half the present one) for the
$\Lambda_{(c)}$ and $\Sigma_{(c)}$  baryons due to the inclusion
of an effective coupling for each baryon species, $C_j(\tilde
\sigma)$,  where ``j" is the label for baryons. This factor was
introduced to  mimic the baryon structure.  To the best of our
knowledge, no temperature effects have been studied within this
framework.

\begin{figure}[t]
\begin{center}
\includegraphics[width=10cm]{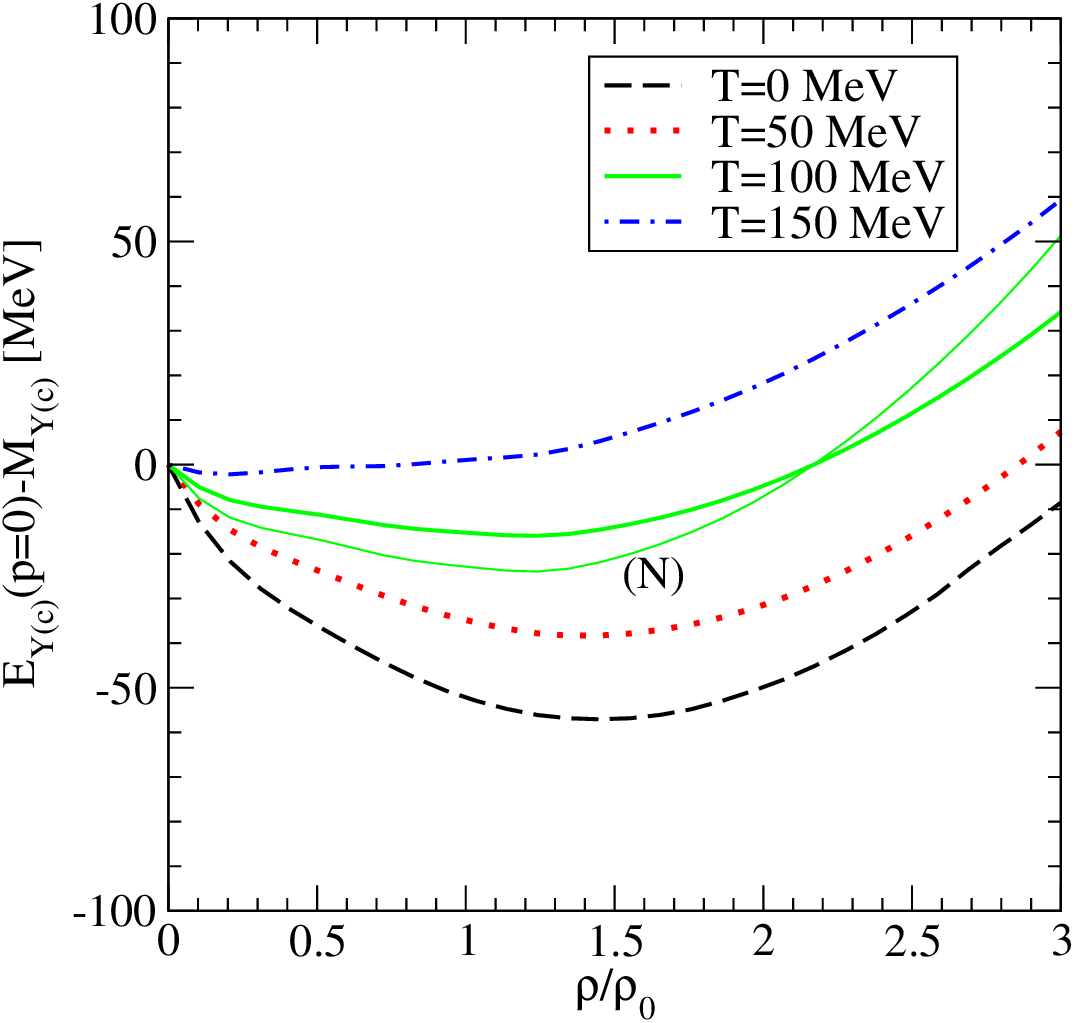}
\caption{The potential at zero momentum for the hyperons ($Y = \Lambda, \Sigma$) and charmed 
baryons ($Y_c= \Lambda_c,
\Sigma_c$) as a function of the density for different temperatures. 
 The thin solid line
 displays the nucleon potential at $T=100$ MeV. }
\label{fig_pot}
\end{center}
\end{figure}

The potential at zero momentum $V=E_{Y(c)}(\vec{p}=0)-M_{Y(c)}$
for different baryon species obtained in the present work is shown
in Fig.~\ref{fig_pot} as a function of the density for different
temperatures. Obviously, the potential for hyperons and charmed
baryons follows the simple light quark counting rule as compared
with the nucleon potential: $V_{Y_{(c)}}=2/3 V_N$. The attraction
at $\rho=\rho_0$ and $T=0$ MeV is about  $-50$ MeV,
the size of which gets reduced as temperature increases turning
even into repulsion, especially at higher densities. This behavior
results from the fact that the temperature independent vector
potential takes over the strongly temperature-dependent scalar
potential which decreases with temperature.

The second medium effect is the Pauli exclusion principle acting on the nucleon
in the  intermediate  $D (\bar D) N$
loop function. This is implemented by replacing the single free-nucleon
propagator in the loop function by the corresponding in-medium one:
\begin{equation}
G_N(p_0,\vec{p},T) =
\frac{1-n(\vec{p},T)}{p_0-E_N(\vec{p},T)+{\rm i}\varepsilon} +
\frac{n(\vec{p},T)}{p_0-E_N(\vec{p},T)-{\rm i}\varepsilon} \ ,
\label{eq:nuc}
\end{equation}
where the effect of the temperature is contained in the nucleon Fermi-Dirac
distributions and single-particle energies.

The third medium effect is the dressing of the mesons, due to their interactions with the surrounding nucleons in the course of propagating through nuclear 
matter.
In particular, we will consider the dressing of pions and $D(\bar D)$ mesons.
The reason for not doing so for other mesons will be stated below.

The meson dressing is represented by the in-medium meson self-energy
$\Pi_i (q_0, \vec{q}, T)$, where $i= D, \bar D, \pi$ in the present case.
This quantity  appears in the corresponding
in-medium single meson propagator:
\begin{equation}
D_i(q_0,\vec{q},T) = \frac{1}{q_0^2-\vec{q}\,^2 - m_i^2 -
\Pi_i(q_0,\vec{q},T)} \ .
\label{eq:prop1}
\end{equation}
In the Lehmann integral representation, the meson propagator may be expressed in terms of the spectral function $S_i(q_0,\vec{q},T)$ as
\begin{equation}
D_i(q_0, \vec{q}, T)= \int^{\infty}_0 \frac{S_i(\omega, \vec{q},
T)}{q_0-\omega+{\rm i}\varepsilon} \, d\omega -\int^{\infty}_0\frac{S_{\bar i}(\omega,
\vec{q}, T)}{q_0+\omega -{\rm i}\varepsilon} \, d\omega ,
\label{eq: prop2}
\end{equation}
where $\bar i$ is the anti-particle to meson $i$. Then we easily relate the self-energy and spectral function as
\begin{equation}
S_i(q_0,{\vec q}, T)= -\frac{1}{\pi} {\rm Im}\, D_i(q_0,{\vec q},T)
= -\frac{1}{\pi}\frac{{\rm Im}\, \Pi_i(q_0,\vec{q},T)}{\mid
q_0^2-\vec{q}\,^2-m_i^2- \Pi_i(q_0,\vec{q},T) \mid^2 } \ .
\label{eq:spec}
\end{equation}

For the case of pions, we incorporate the self-energy at finite
temperature given in the Appendix of Ref.~\cite{Tolos02}, which was
obtained by incorporating the thermal effects to the $T=0$ pion
self-energy model given, for instance, in
Refs.~\cite{Oset90,Ramos94}. We recall that the pion self-energy
in nuclear matter at $T=0$  was obtained by adding to a small
repulsive and constant $s$-wave part \cite{Sek83-Mei89}, the
$p$-wave contribution coming from the coupling to 1$p$-1$h$,
1$\Delta$-1$h$ and 2$p$-2$h$ excitations together with short-range
correlations. These correlations are mimicked by the Landau-Migdal
parameter $g'$, taken from the particle-hole interaction described
in Ref.~\cite{Oset82}, which includes $\pi$ and $\rho$ exchanges
modulated by the effect of nuclear short-range correlations.

In the case of the  $D(\bar D)$ mesons, the self-energy is obtained
self-consistently from the $s$-wave contribution to the in-medium  $D(\bar D)$N amplitude, as will be shown explicitly at the end of this section.

With these medium modifications the propagator loop functions are obtained by four-momentum convolution of meson and baryon
single-particle propagators:
\begin{eqnarray}
{G}_{D(\bar D) N}(P_0,\vec{P},T)= &&\int \frac{d^3 q}{(2 \pi)^3}
\frac{M_N}{E_N (\vec{P}-\vec{q},T)} \times  \nonumber \\
&&\left[ \int_0^\infty d\omega
 S_{D(\bar D)}(\omega,{\vec q},T)
\frac{1-n(\vec{P}-\vec{q},T)}{P_0- \omega - E_N
(\vec{P}-\vec{q},T)
+ {\rm i} \varepsilon} \right. \nonumber \\
&&+ \left. \int_0^\infty d\omega
 S_{\bar D (D)}(\omega,{\vec q},T)
\frac{n(\vec{P}-\vec{q},T)} {P_0+ \omega -
E_N(\vec{P}-\vec{q},T)-{\rm i} \varepsilon } \right] \ ,
\label{eq:gmed}
\end{eqnarray}
for $D(\bar D)N$ states and
\begin{eqnarray}
{G}_{\pi Y_c}(P_0,\vec{P}, T)&= & \int \frac{d^3 q}{(2 \pi)^3} \frac{M_{Y_c}}{E_{Y_c}
(\vec{P}-\vec{q},T)} \int_0^\infty d\omega
 S_\pi(\omega,{\vec q},T) \nonumber
\\
& \times & \frac{1+n_{\pi}({\vec q},T)}{P_0- \omega - E_{Y_c}
(\vec{P}-\vec{q},T) + {\rm i} \varepsilon}  \ , \label{eq:gmedpion}
\end{eqnarray}
for $\pi \Lambda_c$ or $\pi \Sigma_c$ states, where $P=(P_0,\vec{P})$ is the
total two-particle four momentum and ${\vec q}$ is the meson momentum in the
nuclear matter rest frame. Note that, for the $DN$ loop function, the
$S_{\bar D}(\omega, \vec q \ )$ spectral function
appearing in the subdominant second term on the r.h.s. of Eq.~(\ref{eq:gmed})
is assumed to be a free-space delta function.
The $\pi Y_{(c)}$ loop function incorporates the
$1+n_{\pi}(\vec{q},T)$ term,  with $n_{\pi}(\vec{q},T)$ being the Bose
distribution of pions at temperature $T$, in order to account for the 
contribution from thermal pions at finite temperature. Note that, by assuming
perfect isospin symmetry, we set the pion chemical potential to zero in
$n_{\pi}(\vec{q}, T)$.

For $\eta(\eta^\prime) Y_c$, $K \Xi_c (\Xi_c^\prime)$ and $D_s Y$ states, the
corresponding meson lines (propagators) are not dressed by  self-energy
insertions.  In the case of the $\eta, \eta^\prime$ mesons, this is a
reasonable approximation, because the coefficients coupling the
$\eta(\eta^\prime) Y_c$ channels with the $DN$ channel are small  (see Tables I
and II in Ref.~\cite{MIZ06}).   Containing an ${\bar s}$-quark, the $K$ couples
weakly to nucleons, and its spectral function may be approximated by  the free
space one, viz. by a delta function. We could include a moderate repulsive
in-medium shift to the kaon mass, consistent with the repulsion predicted by a
$T\rho$ approximation or more sophisticated models \cite{pentaquark}, but our results are insensitive to this shift due
to the zero couplings of these channels to $DN$. As for the spectral function
of the $D_s^+$ meson appearing in the in-medium $D_sY$ channels, it has been
shown \cite{LUT06} that, in addition to the quasi-particle peak, it presents a
lower energy mode associated with an exotic resonance predicted around 75 MeV
below the $D_s^+ N$ threshold \cite{HOF05}. Therefore, with large coupling
coefficients for transitions $DN \leftrightarrow D_s Y$, one may eventually
have to solve an extended in-medium self-consistent coupled-channel problem
combining the $C=1, S=0$ ($DN$) and  $C=1, S=1$ ($D_s N$) sectors. Work along
this line is in progress.

Lastly, we state that the in-medium $D(\bar D)$ self energy is obtained
by integrating  $T_{D (\bar D)  N}$ over
the Fermi distribution for nucleon momentum at a given temperature as
\begin{eqnarray}
\Pi_{D(\bar D)}(q_0,{\vec q},T)= \int \frac{d^3p}{(2\pi)^3}\, n(\vec{p},T) \,
[{T}^{(I=0)}_{D(\bar D)N}(P_0,\vec{P},T) +
3{T}^{(I=1)}_{D(\bar D)N}(P_0,\vec{P},T)]\ , \label{eq:selfd}
\end{eqnarray}
where $P_0=q_0+E_N(\vec{p},T)$ and $\vec{P}=\vec{q}+\vec{p}$ are
the total energy and momentum of the $D(\bar D)N$ pair in the nuclear
matter rest frame and the values ($q_0$,$\vec{q}\,$) stand  for
the energy and momentum of the $D(\bar D)$ meson also in this
frame. Recall that $\Pi_{D(\bar D)}(q_0, \vec q, T)$ must be
determined self-consistently since it is obtained from the
in-medium amplitude $ T_{D(\bar D)N}$ which contains the $D(\bar
D)N$ loop function $G_{D(\bar D)N}$, and this last quantity itself
is a function of $\Pi_{D(\bar D)}(q_0, \vec q, T)$. From  this we
obtain the corresponding spectral function to complete the
integral for the loop function $G_{D(\bar D) N}(P_0, \vec{P}, T)$
as in Eq.~(\ref{eq:gmed}).

\section{Results and Discussion}
\label{sec:Resul}

\subsection{The $D$ meson spectral function in a hot nuclear medium}

\begin{figure}[t]
\begin{center}
\includegraphics[width=14cm]{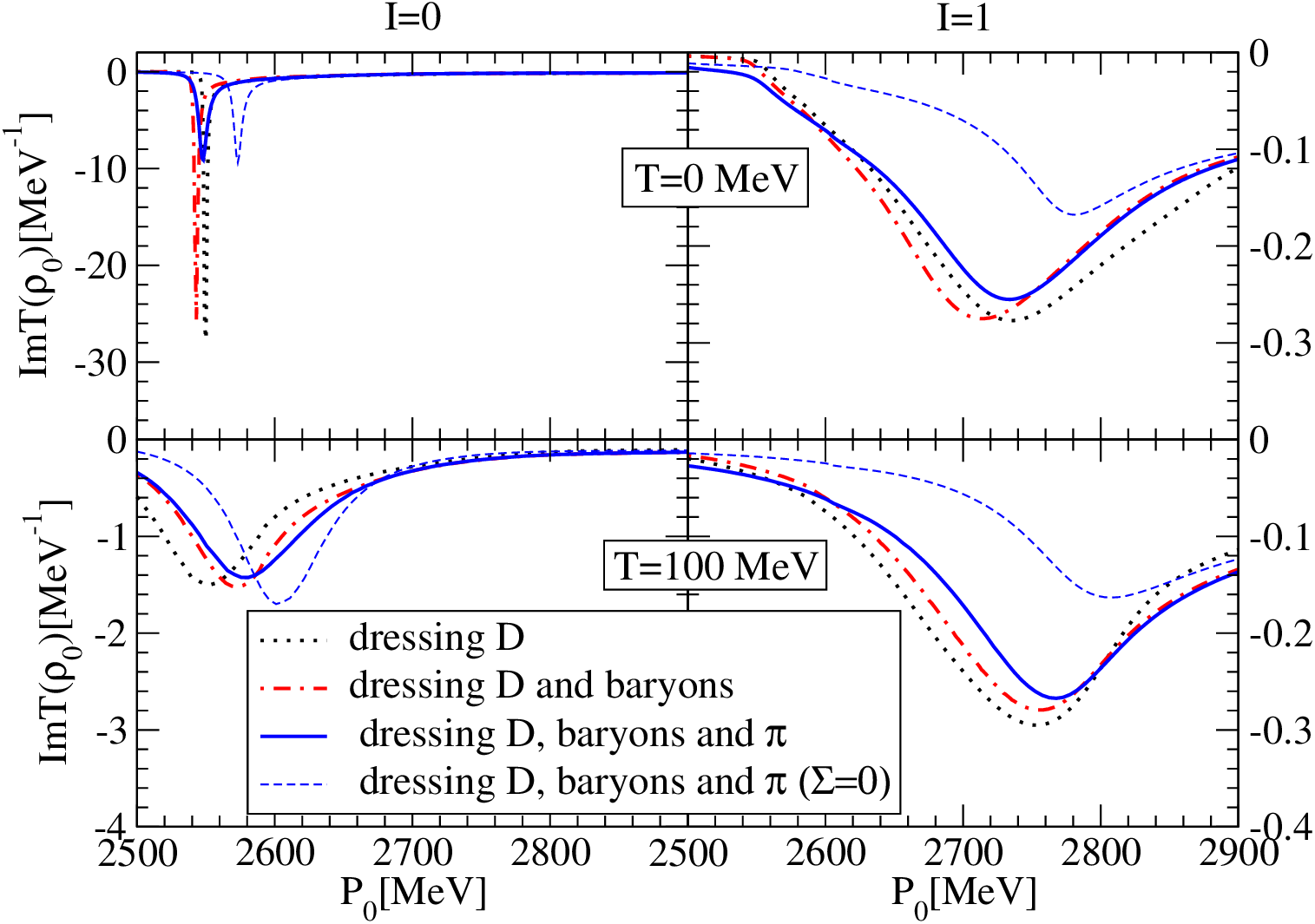}
\caption{Imaginary part of the in-medium interaction for $I=0$ and $I=1$ at
$\rho_0$ as a function of the center-of-mass energy $P_0$ for $T=0$ MeV and
$T=100$ MeV, and for three different approaches in the self-consistent
calculation of the $D$ meson self-energy: i) including only the
self-consistent dressing of the $D$ meson, ii) adding the binding of the 
different baryons in the intermediate states and iii) 
including the baryon binding effects and the pion
self-energy.}
\label{fig_amp}
\end{center}
\end{figure}

We start this section by looking at the in-medium behavior of the $I=0$
$\Lambda_c$ and $I=1$ $\Sigma_c$ resonances, which in the full model of
Ref.~\cite{MIZ06}, appear  at 2593 MeV and 2770 MeV respectively, in 
free space. This is summarized
in Fig.~\ref{fig_amp} by the imaginary part of the in-medium $T(DN
\to DN)$ matrix at saturation density $\rho_0=0.17 \ {\rm fm}^{-3}$ for $I=0$ (left column) and $I=1$ (right column)
as a function of the center of mass energy $P_0$ for temperatures $T=0$ (first
row)  and  $T=100$ MeV (second row), respectively. For each of
the four figures, three different lines represent self-consistent
calculations with
increasing sophistications, viz. 
(i) the self-consistent dressing of $D$ mesons only (dotted lines), 
(ii) $D$ meson dressing with the inclusion of {\it mean-field
binding effect} (MFB) on baryons in the loop functions (dash-dotted lines),
and (iii) $D$ meson dressing with the inclusion of both baryon binding effects 
and {\it pion dressing} (PD) in the loop functions (solid lines).
In the figures, the thick lines correspond to model A (viz.
$\Sigma_{DN}\ne 0$) while the thin dashed lines refer to the result only
for Case (iii) within model B ($\Sigma_{DN}=0$). Recall that our
principal interest is in model A.  Comparison of those two model
results will be found later in this section.

Our discussion is first for the zero temperature ($T=0$) case. We
begin by comparing the two first cases.  We recall that
medium effects (excluding baryon binding potentials) lowered the
position of the $\Lambda_c$ and $\Sigma_c$ resonances with respect
to their free-space values \cite{MIZ06}. When baryon binding effects 
are incorporated, these
resonances get even more lowered, as can be seen by comparing the
dotted and dash-dotted lines in Fig.~\ref{fig_amp}. If we had a
mere attractive shift of the nucleon mass of around 75 MeV (the
value of the optical potential at zero momentum) in a $DN$
single-channel (non-self-consistent) calculation, we would be
expecting the same shift in the corresponding MFB amplitude. The
fact that the attractive shifts induced by MFB effects
are of only 6 MeV and 22 MeV for the in-medium $\Lambda_c$
resonance (hereafter denoted as $\tilde\Lambda_c$) and $\Sigma_c$
(denoted as $\tilde\Sigma_c$) respectively, indicates that
coupled-channel effects, momentum dependence of the binding
potentials, and self-consistency play crucial roles in the
determination of the in-medium $DN$ amplitudes.

The widths of both resonances differ according to the phase space available.
For the $I=0$ sector, the lowest threshold in free space is from the $\pi
\Sigma_c$ channel which lies slightly below the $\Lambda_c(2593)$ resonance
position, and constrains this resonance to be narrow. This is also the case in
nuclear matter.  This narrowness is somewhat relaxed by the processes  $\tilde
\Lambda_c N \rightarrow \pi N \Lambda_c, \pi N \Sigma_c$, which open up in medium through the $D$ meson self-energy.
The $I=1$ $\tilde \Sigma_c$ resonance,
which shows a free-space width of 30 MeV \cite{MIZ06},
develops a large width
of the order of 200 MeV, due
to the opening of new absorption processes of the type $\tilde \Sigma_c N
\rightarrow \pi N \Lambda_c, \pi N \Sigma_c$, which is similar to
the case of the  $\tilde \Lambda_c$
but has  a much larger decaying phase space.

As for the effect of pion dressing (PD) implemented in Case (iii), we expected
it to be of minor importance in  the present approach, in contrast
to the findings in \cite{RAM00,Tolos02} for $\bar K N$ amplitude in nuclear matter, and
also in  Ref.~\cite{TOL04,TOL06}  for  in-medium $DN$. This is due to the reduction
factor $\kappa_c \approx 1/4$ in the $DN \leftrightarrow \pi Y_c$  transition (see
Eq.(\ref{eq:TWint}))
 due to the  charm transfer as shown in Ref.~\cite{MIZ06}. Still a small effect is seen in the positions and widths
of  the $\tilde Y_c \  (=\tilde \Lambda_c, \ \tilde \Sigma_c)$ through the
absorption of these resonances by one and two nucleon processes ($\tilde Y_c N
\to Y_c N$ and $\tilde Y_c NN \to Y_c NN$), which open up through
the $1p-1h$ and $2p-2h$ components of the pion self-energy.

With regard to the results for models A and B, we observe that
both models are qualitatively similar. However, the  absence of
the $\Sigma_{DN}$ term in model B produces in-medium resonances at
higher energies, therefore the corresponding widths are larger due
to the increased decaying phase space. As compared with their free-space resonance positions, the $\tilde \Lambda_c$ lies 45 MeV lower and the $\tilde\Sigma_c$ is at  40 MeV below the free value for model A. In model B, the $\tilde \Lambda_c$ lies 20 MeV lower while the $\tilde\Sigma_c$  moves up roughly by 10 MeV. These results are consistent with those obtained in Ref.~\cite{MIZ06} where
mean-field baryon binding is absent. In
the case of the $I=1$ amplitude, the attraction provided by a
self-consistent calculation using model B is not enough to fully
compensate the repulsion induced by Pauli blocking effects.

Now we come to the finite temperature case. The overall effects
from $T \ne 0$ result is the reduction of the Pauli blocking
factor since the Fermi surface is smeared out with temperature.
Therefore, both resonances move up in energy to get closer to
their position in free space while they are smoothen out. The
inclusion  of MFB (viz. Case (ii)) for $T=100$  MeV induces a
shift of both resonances to higher energies, opposite to what is
found for $T=0$,  hence it appears counter intuitive. The reason
behind this is that the potential of each baryon from MFB becomes
less attractive with increasing temperature and also with
increasing momentum (see Fig.~\ref{fig_pot} for evolution in temperature
of the baryonic potentials at zero momentum). At a high enough
value of temperature, the single-particle potential may become
repulsive already at a relatively low momentum. It will then become
more difficult to excite intermediate states if they carry a
repulsive potential and, consequently, the resonance will be
generated at higher energies than in the absence of MFB effects, as is already the
case at $T=100$ MeV and $\rho=\rho_0$. Again, the 
PD does not drastically alter the  resonance positions. At
$T=100$ MeV, $\tilde \Lambda_c$ is  at 2579 MeV for model A and
$\tilde \Sigma_c$ at 2767 MeV, while model B generates both
resonances  at higher energies: $\tilde \Lambda_c$ at 2602 MeV and
$\tilde \Sigma_c$ at 2807 MeV. The spreading of the resonant
structures in $T_{DN}$ with increasing temperature has an
important bearing in the temperature dependence of the $D$ meson
spectral function as we shall discuss below.

\begin{figure}[t]
\begin{center}
\includegraphics[width=14cm]{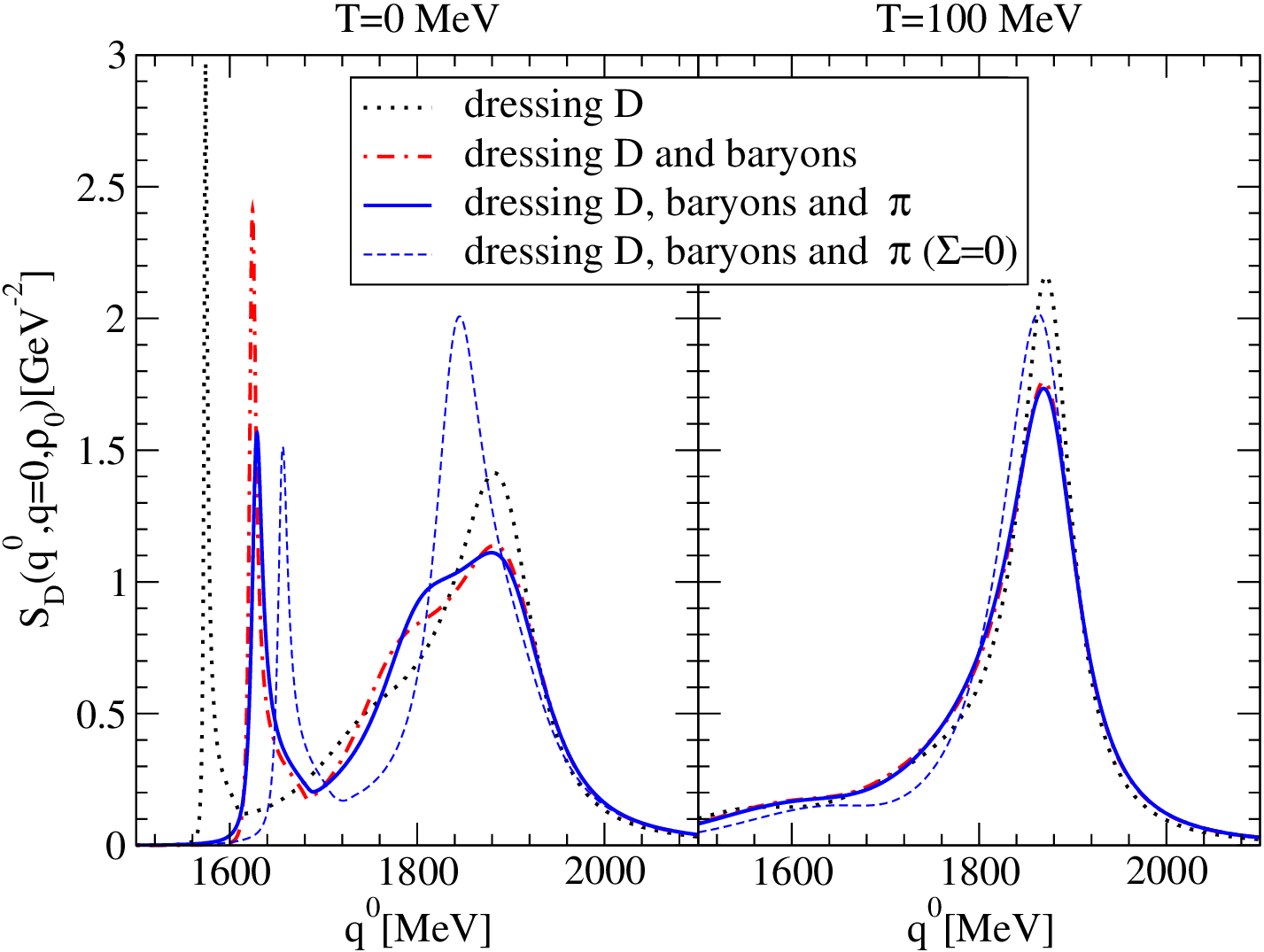}
\caption{The zero-momentum $D$ meson spectral function at $\rho=\rho_0$ for $T=0$ MeV and $T=100$ MeV as a function of
the $D$ meson energy for the previous approaches.}
\label{fig_spec}
\end{center}
\end{figure}

In Fig.~\ref{fig_spec} we display the $D$ meson spectral function
at zero momentum and normal saturation density $\rho_0$ for two
distinct values of temperature: $T=0$ and $T=100$ MeV, and  for
Cases (i) to (iii) (thick lines) for model A. As in the previous
figure, we only show the result from Case (iii) for model B with
thin-dashed lines.

At $T=0$ the spectral function presents two peaks: the one at lower energy is
built up from the $\tilde \Lambda_c N^{-1}$ excitation,  whereas the second one
at higher energy is mainly driven by the quasi(D)-particle  peak but mixes
considerably with  the $\tilde \Sigma_c N^{-1}$ state.  The quasiparticle
energy $E_{qp}(\vec{q}\,)$ may be found from the solution of Re $\left[
D_D(q_0, \vec{q}, T)\right]^{-1}=0$ for $q_0$, hence
\begin{equation}
E_{qp}(\vec{q}\,)^2=\vec{q}\,^2+m_D^2+{\rm
Re}\,\Pi_D(E_{qp}(\vec{q}\,),\vec{q}\,) \ . \label{eq:Qparticle}
\end{equation}
We observe that, once MFB is included (Case (ii)), the lower peak
in the spectral function due to the $\tilde \Lambda_c N^{-1}$ mode goes
up by about $50$ MeV  relative to the Case (i) result. This could
be understood in the following manner: as seen in
Fig.~\ref{fig_amp}, the $\tilde \Lambda_c$ resonance moves to
lower energies by about $6$ MeV upon going from Case (i) to Case
(ii), but at the same time the nucleon energy goes down due to
MFB, hence the peak in the $D$ meson spectral function goes up as
the  $\tilde\Lambda_c N^{-1}$ excitation effectively costs more
energy. In other words, the meson requires to carry more energy to
compensate for the attraction felt by the nucleon.
 The same characteristic feature is  seen also for the $\tilde \Sigma_c
N^{-1}$ configuration that mixes with the quasiparticle peak. Just
in line with the in-medium $T_{DN}$ amplitude studied earlier
(Fig.~\ref{fig_amp}), the PD installed in Case
(iii) does not alter much the position of $\tilde \Lambda_c
N^{-1}$ excitation or the quasiparticle peak. From
Eq.~(\ref{eq:Qparticle}), the corresponding quasiparticle energy
is found at 1855 MeV, i.e. lower than the free mass by 12 MeV.
However, the actual peak appears slightly shifted upwards due to
the energy dependence of the imaginary part of the $D$ meson
self-energy affected by the $\tilde \Sigma_c N^{-1}$
configuration.  For model B (Case iii only), the absence of the
$\Sigma_{DN}$ term moves the $\tilde \Lambda_c N^{-1}$ excitation
closer to the  quasiparticle peak, while the latter fully mixes
with the $\tilde \Sigma_c N^{-1}$ excitation.

When the finite temperature effects are included (see the
right-hand side of Fig.~\ref{fig_spec} for $T=100$ MeV),  the
quasiparticle peak of the spectral function at zero momentum is
found to move closer to the free-space mass value  due to the smearing of the
nuclear matter Fermi surface. The reason is that the self-energy
receives contributions from higher momentum $DN$ pairs that feel a
weaker interaction. Furthermore, structures from the $\tilde Y_c
N^{-1}$ modes seen at $T=0$ are smeared out with increasing temperature,
an effect that was reported earlier in Ref.~\cite{TOL06}. Eventually, at
$T=100$ MeV we are left with a quasiparticle peak at 1869 MeV
(model A) and 1863 MeV (model B), amazingly close to the free
space mass, $M(D)=1867$ MeV, but with a large width due to
collisional broadening. Again, due to the energy dependence of the
self-energy, the positions of these peaks differ slightly from the
value of the quasiparticle energies of 1864 MeV and 1861 MeV,
respectively. The slow fall-off on the left-hand side of the
quasiparticle peak corresponds mostly to the diluted $\tilde
\Lambda_c N^{-1}$ configuration.

\begin{figure}[t]
\begin{center}
\includegraphics[width=14cm]{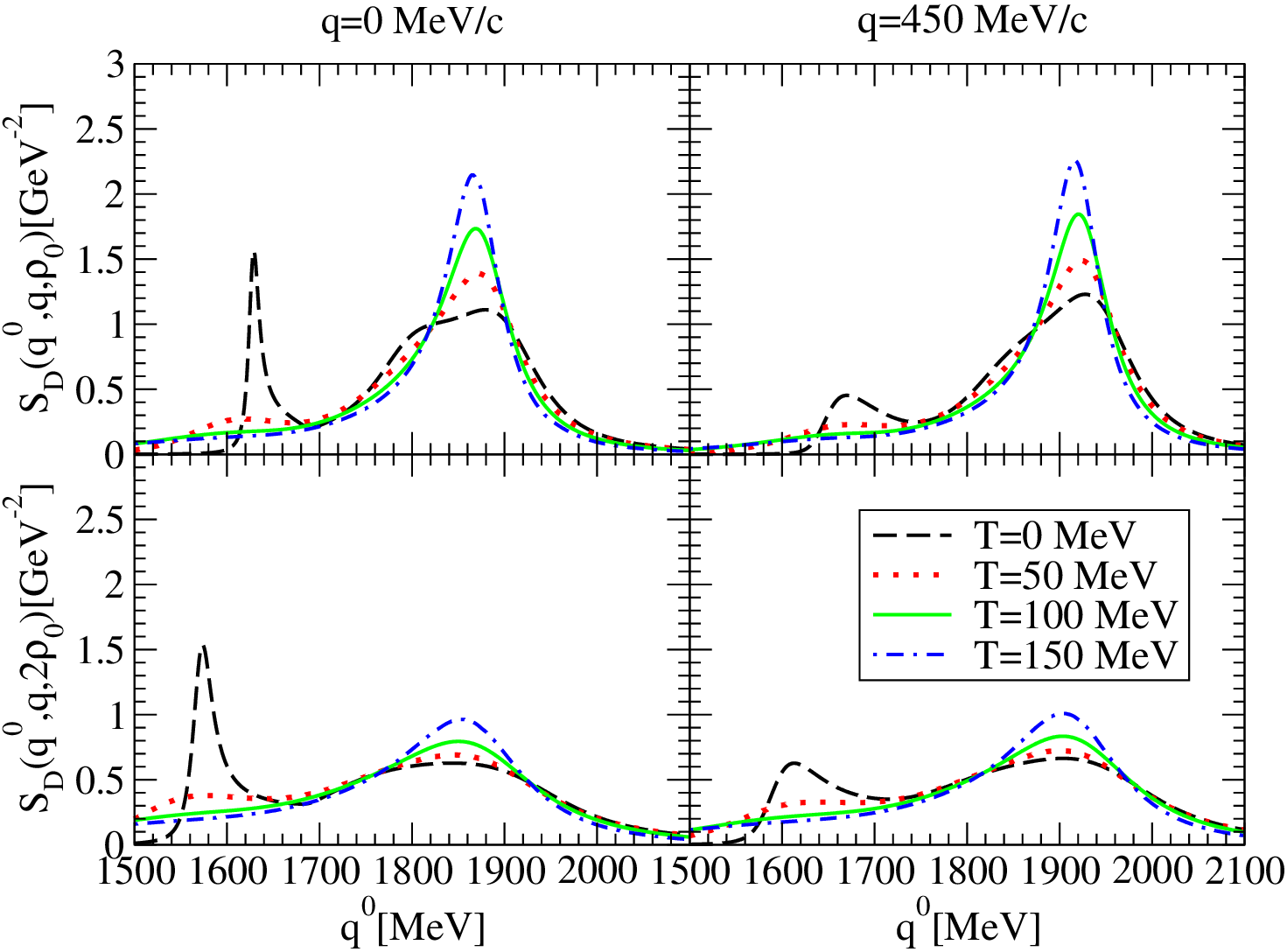}
\caption{The $D$ meson spectral function for $q=0$ MeV/c and $q=450$ MeV/c at $\rho_0$ and $2\rho_0$  as a function of
the $D$ meson energy for different temperatures and for the self-consistent
calculation including the dressing of baryons and pions in model A ($\Sigma_{DN} \ne 0$).}
\label{fig_spectot}
\end{center}
\end{figure}

The evolution of the spectral function as a function of
temperature is presented in Fig.~\ref{fig_spectot} for two
different densities,  $\rho_0$ and $2\rho_0$, and two momenta,
$q=0$ MeV/c and $q=450$ MeV/c, in the full self-consistent
calculation (Case (iii)) for model A.  As already mentioned
before, we observe the dilution of the $\tilde \Lambda_c N^{-1}$
and $\tilde \Sigma_c N^{-1}$  structures with increasing
temperature, while the quasiparticle peak gets closer to its free
value and it becomes narrower.
The widening of the quasiparticle peak for larger nuclear density
may be understood as due to the enhancement of collision and absorption processes.
As a result, the quasiparticle peak position  is difficult to
extract directly from the plot at high densities. As for the
structure in the lower values of $q_0$ due to the $\tilde
\Lambda_c N^{-1}$ configuration, it moves down with increasing
nuclear matter density  due to the lowering in the position of the
$\tilde\Lambda_c$ resonance induced by the more attractive
$\Sigma_{DN}$ term, as based on our experience in
Ref.~\cite{MIZ06}. Then by picking up the case with $q=0,
\rho=\rho_0$ (upper left-hand panel in Fig.~\ref{fig_spectot}) as
an example, we want to analyze the behavior of the same
structure as a function of $T$.  This $\tilde \Lambda_c N^{-1}$
particle-hole configuration  evolves from  a sharp/narrow peak
with very little strength below $q^0=1600$ MeV for $T=0$ to a
more diffused form at higher temperatures where it extends even
below $q_0=1500$ MeV. First, we see from the left-hand panel of
Fig.~\ref{fig_amp} that the in-medium $I=0$ $\tilde \Lambda_c$
resonance becomes broadened and shifted to higher energies with
increasing temperature. We note, however, that a single nucleon-hole state may be created at a
higher energy for higher $T$ because the nuclear matter Fermi surface gets
more diffused. In
addition, at higher $T$, the nucleon single-particle potential
becomes repulsive already at relatively low momenta.
  Hence, the  resulting $\tilde
\Lambda_c N^{-1}$ configuration which dictates the $D$ spectral strength may
spread eventually to lower energies as well  with increasing temperature. So
from what we see in Fig.~\ref{fig_spectot}, along with the wide collision
broadening of the quasiparticle  structure of $D$, this particle-hole
structure  as seen in the $D$ spectral function might well kinematically
facilitate the decay process $J/\Psi N\to \tilde\Lambda_c \bar D$
in a  dense nuclear matter
at finite temperature, say, in  high energy heavy-ion collisions.  This should
of course depend on how the $\bar D$ meson may behave in the same nuclear
matter environment, which we are studying below.

\subsection{$\bar D$ meson in nuclear  matter}

With our models A and B introduced in
Sect.~\ref{sec:Form}, we are able to study the properties of the
$\bar D$ meson in a hot and dense nuclear matter. In fact, as
found in Subsect.~2.2, this case is far easier to deal with
because the $\bar DN$ equation is a single-channel one for both
$I=0$ and $1$ isospin channels. Furthermore, the 
T-W vector interaction in the $I=0$ channel has a strictly {\it
vanishing} interaction strength, so the model B interaction
produces no contribution in this isospin channel. In other words,
the non-vanishing $I=0$ channel contribution in the $\bar DN$
channel comes entirely from the spin-isospin singlet $\Sigma_{DN}$
contribution in model A.  The only available work to date on the
$\bar D$ in nuclear matter is found in Ref.~\cite{LUT06} for the
case of zero ($T=0$) temperature, with which we may compare our
results.

We first present in Table \ref{table:scat} results for the
effective $\bar D N$ interaction in free space.  In particular, we
show the $I=0$ and $I=1$ scattering lengths defined as
\begin{equation}
a_{\bar D N}=-\frac{1}{4\pi}\frac{M_{\bar D N}}{\sqrt{s}}T_{\bar D N \rightarrow \bar D N}
\end{equation}
at $\bar DN$ threshold, where $M_{\bar D N}$ is the total mass of
the $\bar DN$ system.  We use three different models which will be
discussed in the following.  The result from model A and B may not
need any special explanation in view of what has been stated so far
in the  present subsection. However, we want to
reiterate that the $I=0$ contribution in model A is entirely from
the $\Sigma_{DN}$ contribution. Upon comparing these results with
those from Ref.~\cite{LUT06}: $a_{I=1}^{LK}=-0.26 \ {\rm fm}$, and
$a_{I=0}^{LK}=-0.16 \ {\rm fm}$, we see immediately that while the
$I=1$ value is very close, there is a disagreement in the $I=0$
scattering length.  In an attempt to clarify this discrepancy we
have adopted a dimensional regularization method (DR) which was used in 
\cite{LUT06}, but with some minor
modification in the subtraction point as well as in the form of
the interaction of \cite{HOF05}, as discussed in \cite{MIZ06}. The
results, presented as the ``DR" entries in Table~\ref{table:scat},
are very close to the cut-off model B values, which is what one
should have anticipated given the fact that the modifications
implemented in Ref.~\cite{MIZ06} alter the original form of the
Hofmann-Lutz T-W interaction \cite{LUT06} and the unitarization of the
amplitude only marginally,
especially in the present $\bar D N$ channel. Note also
that in our DR approach
the value of the meson decay constant has been chosen to be
$f=f_{\pi}$ as compared with $1.15 f_{\pi}$ in models A and B.
The $I=0$ scattering
length turns out as, of course, zero, with the vanishing
interaction strength. So the $a_{I=0}^{LK}$ value quoted in
Ref.~\cite{LUT06} remains to be somewhat puzzling to us.

 \begin{table}[tb]
    \centering
    \caption{$\bar D N$ scattering lengths ({\rm fm})}
   \begin{tabular}{l | ccc}
     & model A & model B & DR \\
\hline
     $I=0$  & 0.61 & 0 & 0  \\
      (Born approx.) & (0.26) & (0) & (0) \\
 \hline
      $I=1$  & -0.26 & -0.29 & -0.24 \\
      (Born approx.) & (-0.61) & (-0.88)  & (-1.16)
    \end{tabular}
    \label{table:scat}
\end{table}

A recent calculation of Haidenbauer and collaborators \cite{HAI07} employs a
meson-exchange approach supplemented by a short-range one-gluon exchange (OGE)
contribution. It presents a similar $I=0$ scattering length
($a_{I=0}^{H}=-0.07\ {\rm fm}$) but the $I=1$ one is repulsive and almost twice
our result ($a_{I=1}^{H}=-0.45 \ {\rm fm}$). Since the OGE mechanism has
no counterpart within our model, the comparison of results
may not be very meaningful at this point. Nevertheless, it is worth noticing
that about half of the repulsive  scattering length $a_{I=1}^{H}$ comes from the 
hadronic meson-exchange contributions, which can be mapped to a certain extent
to the T-W interaction used in the present work.

To see whether the interaction is reasonably weak, hence the Born
approximation be appropriate or not, we have also calculated the
scattering lengths in that approximation as shown in the table. We
find a big discrepancy between the exact and approximate
values, thus concluding that one has to sum up the whole iterative
series even for this apparently smooth $\bar D N$ interaction as 
noted also in Ref.~\cite{HAI07}.
Thus the Born approximation is not adequate in studying
the in-medium $\bar D$ either. Note that, in the Born
approximation, the $I=1$ scattering length is less repulsive in
model A due to the attractive contribution of the spin-isospin
singlet $\Sigma_{DN}$ term, while the scattering lengths
calculated from the fully iterated amplitude are very similar in
both models. Taking the isospin-averaged scattering length
from the results in Table \ref{table:scat}, one establishes a
repulsive nature for the $\bar D
N$ interaction, even in model A which contains
the attractive effect of the $\Sigma_{DN}$ term.

\begin{figure}[t]
\begin{center}
\includegraphics[width=10cm]{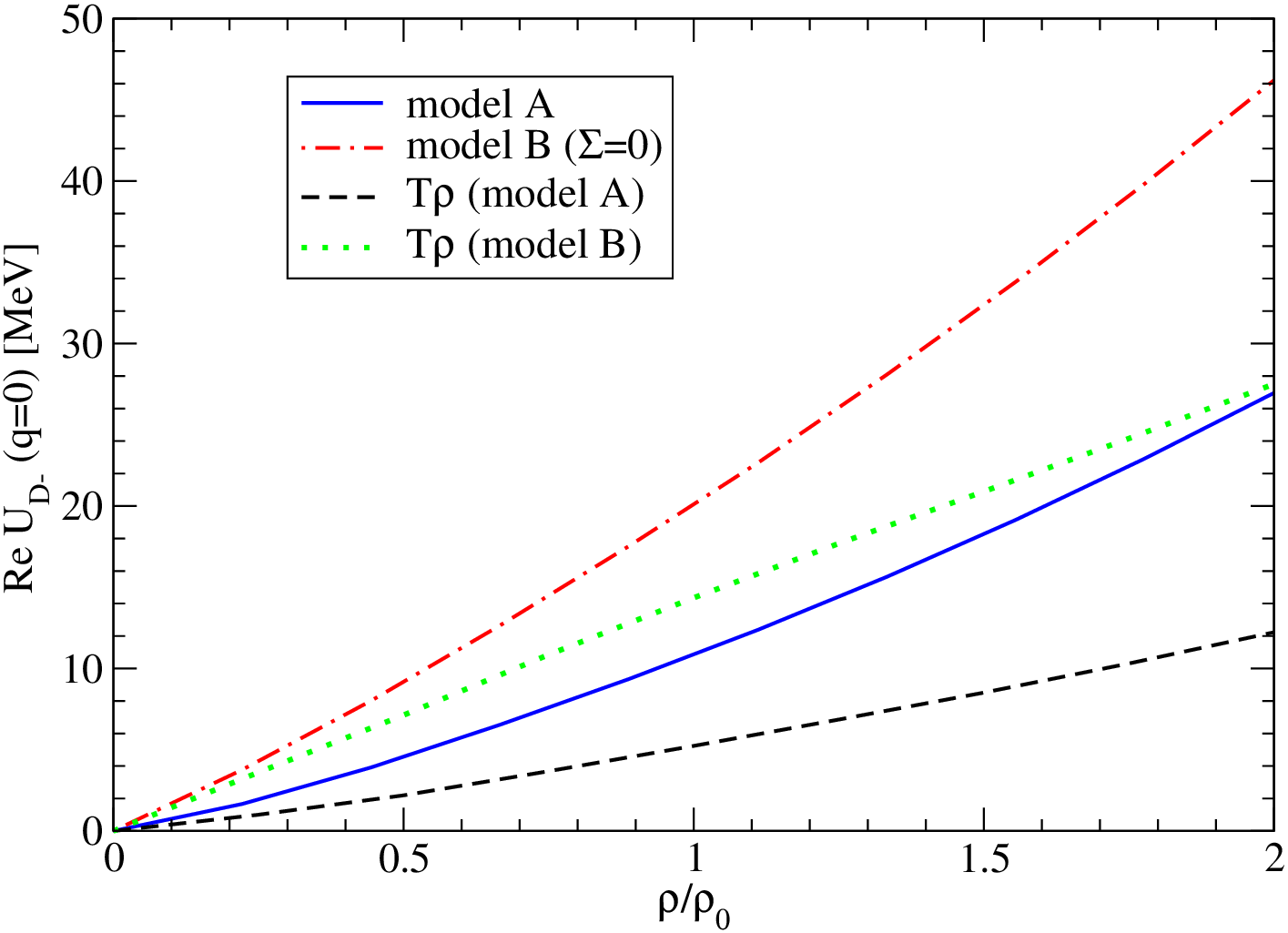}
\caption{The $\bar D$ mass shift in model A  and B ($\Sigma_{DN}=0$) 
including the MFB for nucleons as well as the low-density approximation as 
a function of density.} 
\label{fig_upotdbar}
\end{center}
\end{figure}

The $\bar D$ optical potential in the nuclear medium may be defined
as,
\begin{equation}
U_{\bar D}(\vec{q}\,)=\frac{\Pi_{\bar D}(E_{qp}(\vec{q}\,),\vec{q}\,)}{2\sqrt{m_{\bar D}^2+\vec{q}\,^2}} \,
\label{eq:Dbarpot}
\end{equation}  
and, at zero momentum, it can be identified as the
in-medium shift of the $\bar D$ meson mass. Our results for the
$\bar D$ mass shift are displayed in Fig.~\ref{fig_upotdbar} in the case
of model A (solid line) and B (dot-dashed line) including the MFB for nucleons. 

The inclusion of an attractive $\Sigma_{DN}$ term in model A gives rise
to a less repulsive mass shift at $\rho=\rho_0$, of 
11 MeV, in contrast to the 20 MeV repulsion found for model B.
The absence of resonant states
close to threshold in this ${\bar D}N$ scattering problem suggests
extending
the validity of the low-density theorem to normal nuclear matter
densities or beyond. However, the low-density or $T\rho$ results,
obtained by replacing the medium-dependent 
amplitude by the
free-space one and displayed by the dashed and dotted lines for 
models A and B,
respectively, deviate quite substantially from the corresponding fully
self-consistent
results at a relatively low value of nuclear matter density. The additional sources of
density dependence present in a full calculation can also be visualized
by the deviation of the solid and dot-dashed lines from a linear
behavior. At normal nuclear matter density, the low-density mass shift
for model B is 15 MeV, while the fully-self-consistent result increases
this value to 20 MeV. The same difference of about 5 MeV between the
mass shifts obtained in both approaches is found for model A, as can
be seen in Fig.~\ref{fig_upotdbar}.
Our mass shift of 20 MeV at
$\rho_0$ for model B is similar to the one
in Ref.~\cite{LUT06} within 10\%.

\begin{figure}[t]
\begin{center}
\includegraphics[width=14cm]{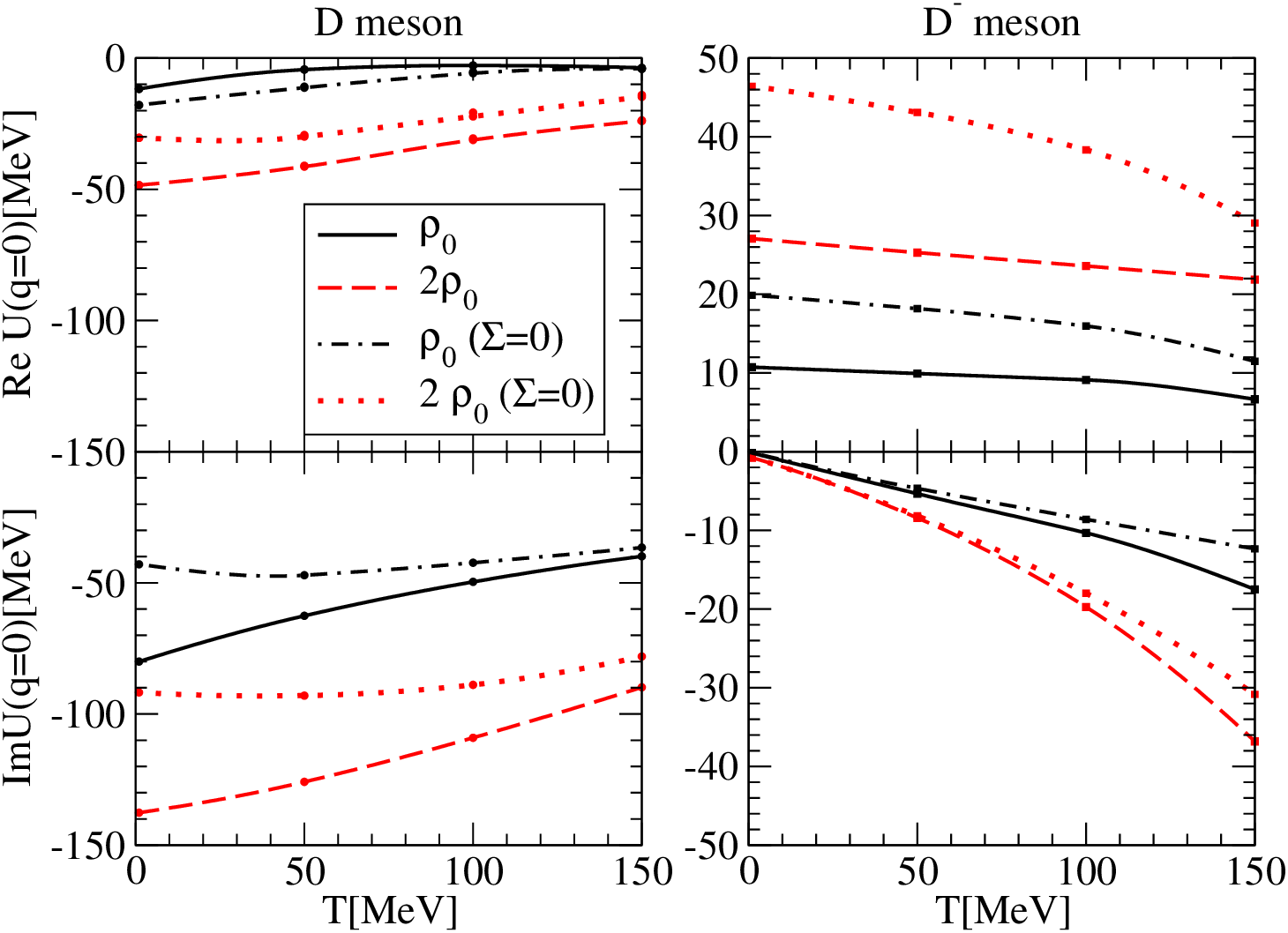}
\caption{The $D$ and $\bar D$ potentials for the full selfconsistent calculation at $q=0$ MeV/c for $\rho_0$ and $2\rho_0$ in models A and B ($\Sigma_{DN}=0$) as a function of temperature.}
\label{fig_upot}
\end{center}
\end{figure}

\subsection{In-medium $D$ and $\bar D$ optical potentials at finite
temperature}

In this last subsection we compare in Fig.~\ref{fig_upot} the $D$ and
$\bar D$ optical potentials at $q=0$ MeV/c as functions of temperature
for two different densities ($\rho_0$ and $2\rho_0$) and for models A
and B.  The $D$ meson potential is calculated self-consistently with
MFB on baryons and with PD, and the $\bar D$ meson is
obtained also in a self-consistent manner only with MFB, since pion
dressing does not enter here. The mass shift for $D$ and $\bar D$
mesons is reduced with temperature since, as observed for the spectral
functions in Fig.~\ref{fig_spectot}, the quasiparticle peak moves
towards the free position. This effect was also observed previously in
Ref.~\cite{TOL06} and it is due to the reduction of the self-energy as
temperature increases because the meson-baryon interaction is averaged
over larger momentum components where it is weaker. For model A (B) and
at $T=0$, we obtain an attractive potential of $-12$ ($-18$) MeV for
$D$ meson while the repulsion for $\bar D$ is $11$ ($20$) MeV. A
similar shift in the mass for $D$ mesons is obtained in
Ref.~\cite{TOL06}. The imaginary part and, hence, the width of the
spectral function for the quasiparticle tends to increase slightly with
temperature for $\bar D$ mesons due to the increase in the collisional
width, while for $D$ mesons it is somewhat reduced. Note, however, the 
different energy scales used in the $D$ and $\bar D$ plots.  In fact,
the situation is more involved for $D$ mesons. On the one hand, the
collisional width due to $DN \to DN$ processes also increases with
temperature, but at low $T$ the $D$ meson width is largely dominated by
the mixing of the quasiparticle peak to the $\tilde\Sigma_c N^{-1}$ components 
of the $D$ meson
self-energy. This is also the reason why the quasiparticle peak is located 
at a lower energy for model B, contrary to what one expects, as explained in the subsequent paragraph  below.
As $T$ increases, the $\tilde\Sigma_c$ resonance gets
diluted and, correspondingly, the width decreases. It is expected, however,
that the width will eventually increase with $T$ when it 
becomes mostly of collisional origin at high enough temperatures.

With regard to the effect of the $\Sigma_{DN}$ term, we find that,
for $\bar D$ mesons, its inclusion substantially reduces the
repulsion, independently of the temperature and density, since
the dominant repulsive $I=1$ scattering length is partly
compensated by the attractive $I=0$ one. However, this simple
picture can not be applied to the $D$ meson due to the presence of
the $I=1$ $\tilde \Sigma_c$ close to the $DN$ threshold. We see that, in this
case, there is a crossover when we go from $\rho_0$ to $2\rho_0$.
The inclusion of the $\Sigma_{DN}$ term alters the position of the
$\tilde \Sigma_c$ close to the $DN$ threshold (as seen in
Fig.~\ref{fig_amp}). Therefore, while it has the expected
attractive effect to the real part of the potential at $2 \rho_0$,
it effectively induces a repulsive effect for $\rho_0$. The
imaginary part increases with the inclusion of $\Sigma_{DN}$ term
for both mesons.

\section{Summary and Conclusions}
\label{sec:Conclusion}

We have performed a hadronic self-consistent
coupled-channel calculation of the $D$ and $\bar D$ self-energies
in symmetric nuclear matter at finite temperature taking an
effective meson-baryon Lagrangian that combines the charmed meson
degree of freedom in a consistent manner with the chiral unitary
models. This interaction consists of a broken  $s$-wave SU(4)
Tomozawa-Weinberg (T-W) contribution supplemented by a
scalar-isoscalar  $\Sigma_{DN}$ term interaction.  The
corresponding in-medium solution at finite temperature obtains the
dressing of $D (\bar D)$ by Pauli blocking effects, dressing of
$\pi$ (PD), and the nuclear mean field binding effect (MFB) not
only on the nucleons, but also on the charmed and strange hyperons
by a finite-temperature $\sigma$-$\omega$ mean-field calculation.

In nuclear matter at $T=0$, the  dynamically generated $I=0$ $\tilde \Lambda_c$
and $I=1$ $\tilde \Sigma_c$ resonances in the $C=1$ charm sector lie around 40
MeV below their free space values. Also at $T=0$, the baryon binding  results
in an attractive mass shift for those resonances as compared to the case with
no such effect. But, as we incorporate finite temperature, those resonances
tend to move back to their free position acquiring a remarkable width due to
the smearing of the Fermi surface.

The $\tilde \Lambda_c$ and $\tilde \Sigma_c$ resonances induce resonant-hole
excitation modes that are clearly seen in the low-temperature $D$ meson
spectral function. The width of the distribution in fact reflects the overlap
of the quasiparticle peak with the $\tilde{\Sigma}_c N^{-1}$ components of the
$D$ meson self-energy. As temperature increases, these modes tend to smear out
and the $D$ meson spectral function becomes a single pronounced  quasiparticle
peak close to the free $D$ meson mass with fairly extended tails, particularly
to the lower energy side of the distribution.  At high temperature the width of
the quasiparticle peak gets reduced slightly, so most of the distribution of 
the spectral function concentrates around the quasiparticle energy, although 
maintaining the overall strength in its lower energy part. As density
increases, the quasiparticle peak broadens and the low-energy strength,
associated to the $\tilde{\Lambda}_c N^{-1}$ components and related to $Y_c \pi
N^{-1}$, $Y_c N N^{-2}$, \dots absorption  modes, obviously increases.

In the $\bar D N$ sector, we have first obtained the free space $I=0$ and $I=1$
scattering lengths. While our repulsive $I=1$ value of $a_{I=1} \sim -0.3$ fm 
is in good agreement with Lutz and Korpa results \cite{LUT06},  the finite
value for the $I=0$ scattering length found in this latter reference is in
contrast to the zero value found here for model B due to the vanishing
$I=0$ coupling coefficient of the corresponding pure Tomozawa-Weinberg  $\bar
DN$ interaction. Our results are, however, consistent
with a recent calculation based on a meson-exchange model 
supplemented by a short-range
one-gluon exchange contribution \cite{HAI07}. 
For model A, we obtain a non-zero value of the $I=0$ scattering length,
dictated entirely by the magnitude of the $\Sigma_{DN}$ term, which takes a rather
conservative value in our present work. We have also observed that, in spite of
the weakness of the ${\bar D}N$ interaction and the absence of resonances close
to threshold, the Born approximation is not sufficient to describe the
free-space ${\bar
D}N$ interaction at low energies. 

As for medium effects, they
induce a repulsive shift
in the $\bar D$ meson mass of 11 MeV (20 MeV) for model A (B) in 
nuclear matter at saturation density. Although the medium modifications
of the ${\bar D}N$
interaction are more moderate than in the case of $DN$, we observe that
the low-density
approximation breaks down at relatively low densities. At nuclear matter
saturation density, the ${\bar D}$ meson mass shifts obtained from
a fully self-consistent calculation are 5 MeV larger than those of 
the low-density approximation.
The temperature dependence of the repulsive real part of the ${\bar D}$ optical
potential is very weak, while the imaginary part increases steadily due to the
increase of collisional width. The picture is somewhat different for the $D$
meson. At low temperature, the corresponding quasiparticle peak is
already quite broad due to the
overlap with the $\tilde{\Sigma}_c N^{-1}$ mode. As temperature increases
the later mode tends to dissolve and, with the overlap being reduced, one 
observes an overall decrease in the width of the distribution in spite of the 
increase of collisional broadening.

Taking into account our results, we might look at the
question of possible $\bar{D}$ bound states once discussed in \cite{TSU99}.
While $D^-$ -mesic nuclei systems will always be bound by the Coulomb interaction,
it would be interesting to see whether strongly bound nuclear states or even
bound $\bar{D}^0$ nuclear systems might exist. From the results of
Fig.~\ref{fig_upotdbar} we see that, even for model A, the
${\bar D}$-nucleus optical potential at zero momentum is repulsive, hence ruling
out this possibility.
However, as mentioned earlier, the $\Sigma_{DN}$ term, which contributes
attractively to both isospin channels of the $\bar DN$
interaction, has been given a conservative value in our present approach.
Although the magnitude of the $\Sigma_{DN}$ term may be 
made larger, we recall that it is not a free parameter but 
is constrained by the coupled $DN$
channel, in particular by the properties of the $\Lambda_c(2593)$ resonance.
Therefore, it will be first necessary to see whether a larger $\Sigma_{DN}$ term
constrained by the properties of the $DN$ interaction may
still produce an attractive isospin averaged interaction, which could  
then even allow for the existence of $\bar D^0$-nucleus bound states. 
In any event, it is clear
that $D^-$-mesic nuclei provide a valuable source of information for 
determining the 
sign and size of the ${\bar D}$ mass shift at subnuclear densities.
An experimental observation of bound $D$ nuclear states is
ruled out by the large width and moderate attraction found for the $D$
meson optical potential.

The other point of interest from the present study is related to the possible
hadronic  mechanism for the suppression of the $J/\Psi$ production in
relativistic heavy-ion collisions. As stated in the introduction, a good part
of the earlier interest was to see if the masses of $D$ and $\bar D$ get
reduced so that the $J/\Psi \to D \bar D$ may proceed {\it
exothermically} \cite{TSU99, SIB99, MIS104, LUT06} in nuclear matter,
hence contributing to the $J/\Psi$ suppression. The difference in the two
thresholds in vacuo for this process is $\Delta E\equiv E_{th}(D+\bar
D)-E_{th}(J/\Psi) \approx 650$ MeV, which should be overcome in one way or
another for this to go spontaneously or at the  cost of small energies. Since
charmonia including the $J/\Psi$ are $c \bar c$ bound states which do not
contain any light quarks, one normally assumes that the medium modification
they might undergo should be minimum, thus their in-medium masses are not
expected to be very different from those in vacuo. In the present work, we have
observed that in-medium $\bar D$ mass increases typically by $10 - 20$ MeV. 
On the other hand, as mentioned in the last section, from
Fig.~\ref{fig_spectot}, the tail of the quasiparticle peak of the $D$ spectral
function extends with a non-negligible strength to lower {\it ``mass"} values due
to the thermally spread $\tilde  Y_c N^{-1}$ particle-hole configurations. 
So one might expect that some spontaneous leakage for $J/\Psi N \to 
{\tilde Y}_c  \bar D \rightarrow  Y_c \pi \bar D$, $J/\Psi N N \to Y_c N  \bar
D, \dots$
might effectively be possible.  However, it looks very unlikely that the lower
part of this spectral function tail extends from the quasiparticle peak as far
down by $600$ MeV while keeping relevant strengths. So the
disappearance of the $J/\Psi$ through such processes, even if helped by the
thermally excited nucleons, will not proceed. These
processes might still go  endothermically through  collisions induced by a surrounding large collection of 
co-moving hadrons for which  the effective reaction threshold may  get
lowered. But its efficiency is very questionable. So the direct disappearance 
of the existing $J/\Psi$ looks improbable.

Then, what about the possibility of reducing the supply of this charmonium from
its excited state partners such as $\chi_{c\ell}(1P),\ (\ell=0,1,2)$ since 
it is
known that an appreciable fraction of $J/\Psi$ production comes from the
radiative decay  of these charmonia \cite{ANT93}?  To look for such a
possibility, we may consider here the following reactions:  $\chi_{c\ell}(1P) N
\to \Lambda_c(2285) \bar D$, $\chi_{c\ell}(1P) N \to \Lambda_c(2285) \pi \bar
D$, $\chi_{c\ell}(1P) N \to \Sigma_c(2445) \bar D$ and $\chi_{c\ell}(1P) N \to
\tilde\Lambda_c(2593) \bar D$. Note that the masses of these $\chi_{c\ell}$'s
are $3415,\ 3511$, and $3556$ MeV in ascending order for $\ell$. Before
proceeding, we note that in free space the first three reactions are already
endothermic for all three $\chi_{c\ell}$'s, while the last one is closed for
the two lowest ones. Upon taking into account the MFB effects studied in the
present work,  it should be safe to speculate that the same first three
reactions do take place in a hot nuclear matter as well. As for the last one,
we might also claim that it could proceed in medium since the $\tilde
\Lambda_c$ will develop a sufficient width as seen, for example, in the
in-medium ${D}N$ amplitudes of Fig.~\ref{fig_amp} or reflected in the
extended low energy tail of the $D$ meson spectral functions in
Fig.~\ref{fig_spectot}.   Therefore part of the feeding of the $J/\Psi$ from
its excited state partners will certainly be reduced. A similar mechanism might well reduce the feeding of $J/\Psi$ from the decay of the $\Psi'$.  The arguments presented
here are only kinematical in nature, so even a simple dynamical model for the
relevant in-medium reactions may need to be employed to support them further.

\section{Acknowledgments}

T.M. is grateful to the support offered through the host (A.R) for his stay in Barcelona, and for
part of his travel. This work is partly supported by the EU contract
FLAVIAnet MRTN-CT-2006-035482,
by the contract FIS2005-03142 from MEC (Spain) and
FEDER and
by the Generalitat de
Catalunya contract 2005SGR-00343. This research is part of the EU
Integrated Infrastructure Initiative Hadron Physics Project under
contract number RII3-CT-2004-506078.
L.T. wishes to acknowledge support from the BMBF project
``Hadronisierung des QGP und dynamik von hadronen mit charm quarks'' 
(ANBest-P and BNBest-BMBF 98/NKBF98).


\end{document}